# Multi-speaker Recognition in Cocktail Party Problem

Yiqian Wang, Wensheng Sun[1]

Beijing University of Posts and Telecommunications, Beijing, China
yqwang_c@163.com, sunws@bupt.edu.cn

**Abstract.** This paper proposes an original statistical decision theory to accomplish a multi-speaker recognition task in cocktail party problem. This theory relies on an assumption that the varied frequencies of speakers obey Gaussian distribution and the relationship of their voiceprints can be represented by Euclidean distance vectors. This paper uses Mel-Frequency Cepstral Coefficients to extract the feature of a voice in judging whether a speaker is included in a multi-speaker environment and distinguish who the speaker should be. Finally, a thirteen-dimension constellation drawing is established by mapping from Manhattan distances of speakers in order to take a thorough consideration about gross influential factors.

## 1  Introduction

Cocktail party problem describes a psycho-acoustic phenomenon due to masking effect. [1] For instance, a person in the noisy environment of a cocktail party can focus on a specific speech of another person while ignoring speeches of the others. The top-down attention of the person can affect the process, which contains two spheres, speaker recognition and voice filtering. [2] This paper chiefly considers and imitates the first part of subconscious thinking process which helps us to locate the sound of a talker. The compound logic in searching out who is the most probable speaker and who is the least possible one is named as statistical decision theory in multi-speaker recognition.

In hypothetical scenery, Mr. Bright wants find out one of his friends in a cocktail party only by auditory sense. Since the timbre of that friend is already known, he needs to match the voice in memory to the mixture of sounds in the party. When some speakers are communicating at the same area, they can be included into a group, thus those people participating in the party can be separated into varied groups. Then the task is simplified into two concrete steps, one is to determine whether a single speech belongs to a group of speakers, the other is to decide who is the owner of the voice if it has met the former requirement.

[1] Wensheng Sun (✉)
Beijing University of Posts and Telecommunications, No.10 Xutucheng Rd. Haidian District, Beijing, China
e-mail: sunws@bupt.edu.cn

In a previous study of speaker recognition, Gaussian mixture models (GMM) and Expectation Maximization (EM) [3] were used to compare two speeches of the same content with slightly distinction in that they were recorded in different time. The algorithm reaches an error rate of $21.1\%$ in contrasting couples of 3.2-second speeches, and the error rate declines when the time of speeches increases. Its text-dependent demerit is ascribed to the limitation of Gaussian mixture models whose model number should be predefined.

Another related study used joint Mel-Frequency Cepstral Coefficients (MFCCs) and Vector Quantization (VQ) algorithm [4] to deal with a similar case, but the content of each speech could be different, making its final result text-independent. The study is completed by comparing each pair of single speeches with silent backgrounds, and it reaches an error rate of $16.6\%$.

Nevertheless, none of them can contrast a single speech with a blend of speeches by varied people. Therefore, this paper used Mel-Frequency Cepstral Coefficients and a statistical decision theory to decide whether a single speech is included in the blend of voices and who is the possessor of the single speech.

Mel-Frequency Cepstral Coefficients can help us to define the acoustics features of a person by emulating the response in human auditory system. [5] The statistical decision theory is our original idea coming up from the book Principles of Communications [6].

In traditional statistical decision theory, the best reception of a digital modulation signal in M notation through an Additive White Gaussian Noise (AWGN) channel is realized by Maximum Posterior-probability (MAP) Algorithm. And it can be simplified into Maximum Likelihood (ML) algorithm or Minimum Euclidean distance algorithm when the prior-probabilities are equal. A digital signal through AWGN channel is comparable to a speech in the multi-talker background, so the best reception rate of the digital signal is parallel to the least judgment error rate in the recognition with background noises.

This paper also builds a thirteen-dimension constellation drawing based on MFCCs, where each spot represents the gross influence of diverse frequencies in a speaker's voiceprint. Making judgment of one-second recordings in three-speaker environment, the error rate is $16.6\%$ if the contents are the same, and $22.6\%$ in text-independent condition.

## 2 Recognition Algorithms

The recognition algorithms combine two parts, a classical feature extraction algorithm MFCCs and a new statistical decision theory proposed in this paper.

## 2.1 Feature Extraction

Mel-Frequency Cepstral Coefficients (MFCCs) are representations of the short-term power spectrum of a sound, based on a linear cosine transform of a log power spectrum on a nonlinear Mel scale of frequency. [7][8] It can concisely imitate the frequency masking effect in human's basement membrane of cochlea, where the lower frequency sounds transmit farther distance and are easier to be recognized than the higher ones.

MFCCs are commonly derived by the following steps: [4]
1) *Frame blocking*
2) *Windowing*
3) *Fast Fourier transform (FFT)*
4) *Mel frequency warping*

Distinct frequencies are perceived non-linearly, so Mel-Scale filter bank can characterize the preciseness of human ear:

$$Mel(f) = 2595 \lg(1 + f/700) \quad (1)$$

5) *Cepstrum*

The MFCCs are resulted from Discrete Cosine Transform (DCT), and $S_k$ represents Mel-scale warping stage:

$$Cn = \sum_{k=1}^{N} \lg S_k \cos\left(\frac{\pi n(k-0.5)}{N}\right) \quad n = 0, 1, \ldots, N-1 \quad (2)$$

After extracting voice features, the coefficients are obtained: thirteen figures for each signal on behalf of varied proportions in different frequency.

These coefficients represent the voiceprint of a person, so the more similar voiceprint people have, the smaller Euclidean distance their Mel-coefficients reach. For instance, $K$ participants have recorded $N$ sentences, and their MFCCs are: $x_1^1 \sim x_{13*N}^1, x_1^2 \sim x_{13*N}^2, \ldots, x_1^{K-1} \sim x_{13*N}^{K-1}, x_1^K \sim x_{13*N}^K$.

In the $i^{th}$ sentence, the Euclidean distance vector of speaker P and speaker Q is: $D = \left\{ \left|x_{1+13*(i-1)}^P - x_{1+13*(i-1)}^Q\right|, \left|x_{2+13*(i-1)}^P - x_{2+13*(i-1)}^Q\right|, \ldots, \left|x_{12+13*(i-1)}^P - x_{12+13*(i-1)}^Q\right|, \left|x_{13*i}^P - x_{13*i}^Q\right| \right\}$.

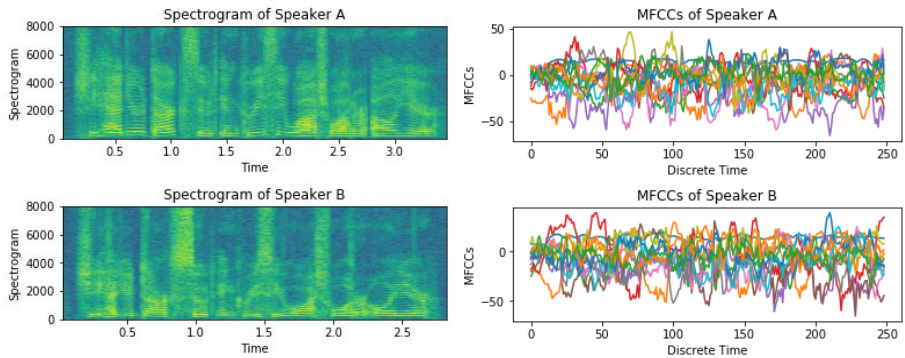

**Fig. 1.** Spectrograms and MFCCs when speaker A and B uttering the same sentence.

The speakers in the recording data come from varied districts in the United States and each of them has recorded ten sentences, including two duplicate sentences and eight distinct ones [9][10][11]. In the recordings of same sentences and distinct sentences, the first part is easier to be recognized, because the features of different speakers have obvious distinctions. The second part is more effortful to be distinguished because the distinctions in content bring interference in extracting the characteristics of multiple talkers.

In the example figure (Figure 1), speaker A is female and speaker B is male. As the graphs show, their voice spectrograms and MFCCs are apparently different, so the voiceprints of them have little similarity.

### 2.2 Statistical decision theory

*1) The initial model*

Taking the simplest case of three-speaker environment into consideration, speaker A, B, and C are talking at the same time, and the acoustic features of them have been acquired by a recorder. Importing a specific voice V, the similarities between V and speaker A, B, and C can be determined by their Euclidean distance vectors. By comparing the Euclidean distance vector between V and each of them, we can figure out which of the speaker V is most likely to be. In addition, we should contrast the characteristics between V and the composition of the three to make sure if V is included.

Supposing the Euclidean distance vector of speaker A and the mixed voice of the three is $D_A$, that of the voice V and the intermixed voice is $D_V$, we can conclude that V has a higher likelihood to be none of them if V has a the closest average Euclidean distance to A but the mathematical mean of $D_V$ is larger than that of $D_A$.

This statistical decision in three-speaker environment can be further improved if we consider the Euclidean distance vector between V and the mixture of two speakers. V is less likely to be the voice of A if it has a nearest Euclidean distance to the blended speech of B and C, so this consideration is a reverse logic to the initial algorithm, and there will be a balance to decide the likability of V and the other three speakers.

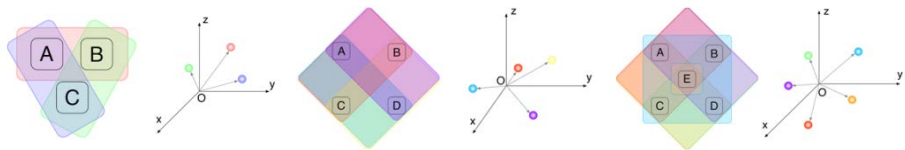

**Fig. 2.** The sketch map of finding the least possible speaker in three to five speakers' environment.

When $N$ speakers are talking at the same time, this algorithm can be popularized from three to $N$ by adding the combinations of more speakers. The examples of different combinations containing two speakers to four speakers are showed in Figure 2.

   *2) The enhanced model*

If we assume that the proportion of every frequency in a speaker's voiceprint follows Gaussian distribution, the distribution of each index in the Euclidean distance vector between V and one of those speakers can be shaped by a Gaussian model.

Extracted from MFCCs, the coefficients vector of V is:
$$x = \{x_1, x_2, x_3, x_4, x_5, x_6, x_7, x_8, x_9, x_{10}, x_{11}, x_{12}, x_{13}\}.$$

The post-probability $P(\omega_i)$ is deduced from Bayesian Theorem:
$$P(\omega_i|x) = \frac{p(x, \omega_i)}{p(x)} = \frac{p(x|\omega_i)P(\omega_i)}{p(x)}, \quad (3)$$

$$P(\omega_i|x) = \frac{p(x, \omega_i)P(\omega_i)}{p(x)} = \frac{p(x|\omega_i)P(\omega_i)}{\Sigma_{j=1}^N p(x|\omega_j)P(\omega_j)}. \quad (4)$$

In the theorem above, $P(\omega_i)$ is the distribution of MFCCs vector $\omega_i$, which represents the voiceprint of A, B, and C when $i = 1, 2, 3$. As is hypothesized above, $P(\omega_i)$ follows Gaussian distribution, $\mu_i$ is mathematical mean of $\omega_i$ and $\sigma_i$ is standard deviation of $\omega_i$, so the distribution of $P(\omega_i)$ is:
$$P(\omega_i) = \frac{1}{\sqrt{2\pi}\sigma_i} \exp\left(\frac{1}{2}\left(\frac{\omega_i - \mu_i}{\sigma_i}\right)^2\right) \quad (5)$$

In the definition, the Manhattan distance of two vectors is the average distance in each dimension:
$$E = \Sigma_{i=1}^N |x_i - \mu_i|. \quad (6)$$

Deduced from (6), the Manhattan distance of two MFCCs vectors is:
$$E_k = \frac{1}{13}\Sigma_{i=1}^{13} \left||(x_i - \mu_i^k)| - \sigma_k\right|, k = 1, 2, 3\ldots \quad (7)$$

Collecting the voiceprint vectors of A, we can conclude them into a constellation drawing with thirteen dimensions, where each spot represents the location of a vector. Because A has recorded ten sentences and each of them may lead to a distinct spot, the area which contains those spots can be enclosed as a circle to represent the characteristic of A. Thus, the area is quite possible to be reached if the input test sentence belongs to A.

Since the spots of a speaker obey normal distribution, it is logical to deduce that $68.27\%$ of each index is included in the interval $\mu_i \pm \sigma_i$, so $99.99\%$ of this area is covered by a new vector $\mu_i \pm \sigma_i$, which is tested to be a felicitous balance to make judgments.

$$P(\mu_i \pm \sigma_i) = \overbrace{\left(\left(\left((0.6827)^{0.6827}\right)^{0.6827}\right)^{\cdots}\right)^{0.6827}}^{13 \text{ times}}$$
$$= 99.99\%$$

Therefore, a sentence can be excluded from belonging to A when the mapping spot of MFCCs vector is too far from that area (imaging that $\mu_i$ is the center of a circle with diameter of $\sigma_i$ in two-dimension constellation drawings, which pile up together to form a thirteen-dimension constellation figure). The following figure (Figure 3) is a visualization of two vectors mapping in thirteen-dimension constellation.

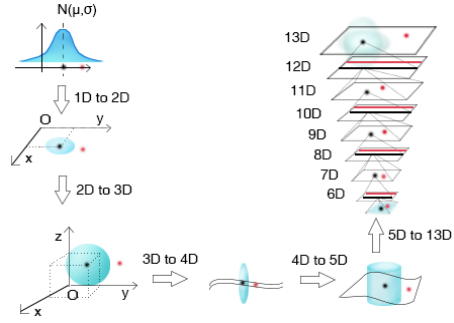

**Fig. 3.** The locations of two MFCCs vectors in 13-dimension model.

*3) The complete theory*

The thorough algorithm in three-speaker environment is as follows:

First, calculate the Euclidean distance vector of V and the mixture of multiple speakers ($D_V$). Then use the training data of each separated speaker to determine an average Euclidean distance vector ($D_E$) likewise. Compare the mathematical mean of $D_V$ and that of $D_E$, and then decide whether V should be included in those talkers.

Second, contrast the MFCCs vector of V and that of the combinations of some speakers, and decide which specific speaker V is less likely to be. If the contrast between V and a combination containing speaker W is higher than that of another one excluding W when the rest speakers are unchanged, we can conclude speaker W is quite improbable to be V.

Third, use $\mu_i \pm \sigma_i$ as a vector and compare it with the MFCCs vector of V when $i$ change from one to $N$ ($N$ is the total number of speakers) in order to find out the Manhattan distances of V and other speakers.

Finally, we can decide who V should be if V has the closest Manhattan distance to a specific speaker and also has the farthest contrast to the mixture without the one.

## 3    Experimental Results and Analysis

In the experiment, input sentences are separated into two parts, the same sentence data and the distinct sentences data. The first part is easier to be distinguished while the second part is more subtle to be decided due to more interference.

By testing numeral combinations of voices in the second step of statistical decision theory above, the recognition error rate approximately goes down in logarithmic form when the number of testing data increases. Moreover, the forms of different combination increases rapidly as the total number of speakers increases, so hardness boosts in making precise decision of its owner with increasing total number.

In three-speaker environment, the final error rate of distinct sentences is $22.6\%$, while that of the same sentences is $16.5\%$. It is quite reasonable because the features of same content sentences are extracted with less interference, making the judgment easier.

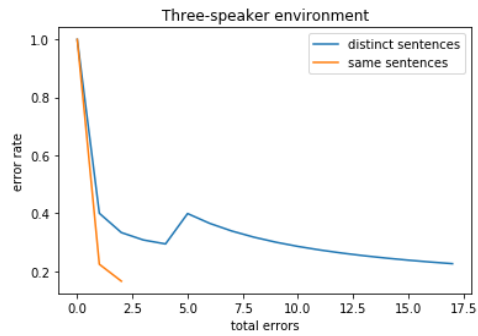

**Fig. 4.** The error rate variation in three-speaker environment.

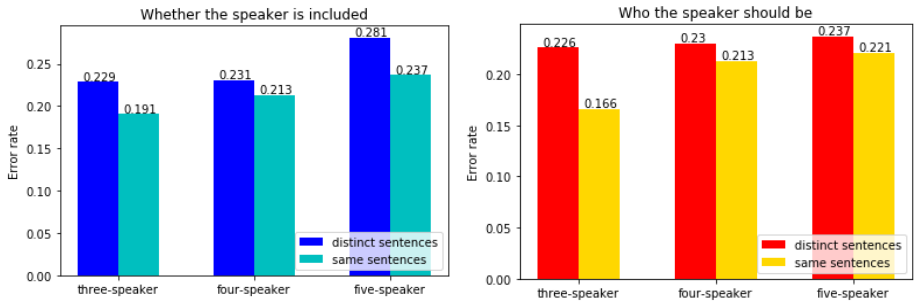

**Fig. 5.** The error rate of whether the speaker is included and who the speaker should be in different kinds of environment.

## 4 Conclusion and Future Development

In this paper, the two prime algorithms are MFCCs and statistical decision theory, and two tasks are resolved: judging whether a voice flow is uttered by one of those speakers and finding the owner of the voice. The error rate in these tasks increases slightly when the total number of speakers rises from three to five.

The uniqueness of this paper is that it can compare a single voice with the blend of varied voices in a moment, and make a synthesized decision of the speaker all at once. But previous works in speaker recognition always make comparisons one by one, and finally obtain the most possible owner of a voice by seeking out the one with the highest possibility,

Since this paper only takes the timbres of people into consideration, its weakness is that judgment error occurs when a person's timbre is so similar to another one that even human's ear cannot correctly distinguish them. If we utilize the accents of people in speeches, the error rate may be narrowed down to some extent. For instance, the accents in differential parts of the United States are varied, some local people may say 'Oh, my gowd' ('Oh, my god') in New Jersey, and 'mah fanger hurts' ('my finger hurts') in Alabama. For those people with similar timbre but different accent, the syllables of their speech can be used in algorithm such as Hidden Markov Model (HMM).

## Acknowledgment

We would like to acknowledge Electronic Information Specialty Group of Universities in Beijing for funding the paper.